# Measure of the hygroscopic expansion of human dentin

H. Gharbi[a], W. Wang[a,b], C. Giraudet[a,c], J.-M. Allain[a,c], *,#, E. Vennat[b,d],#

a. Laboratoire de Mécanique des Solides, CNRS, Ecole polytechnique, Institut polytechnique de Paris, 91120 Palaiseau, France
b. Université Paris-Saclay, CentraleSupélec, CNRS, Laboratoire de Mécanique Paris-Saclay, 91190 Gif-sur-Yvette, France
c. Inria, Inria Saclay-Ile-de-France, 91120 Palaiseau, France
d. URB2i, Université Paris Descartes, Montrouge, France

**\*, corresponding author:**
E-mail address: jean-marc.allain@polytechnique.edu
Phone: +33172925940

#: these two authors contributed equally to the work.

**Background**: Direct dental restoration implies a drying of the dentin substrate. This drying may induce significant strain in the dentin, affecting the bonding efficiency of the restoration. **Objective**: We measure the dilatation of dentine under changes of relative humidity as well as the impact of humidity on dentin elastic properties. This investigates the role of relative humidity variation during dental surgery on restoration lifetime. **Methods**: We have coupled an environmental chamber to control both temperature and humidity on the sample, with an optical microscope to measure precisely the strain on the sample surface, after a quantification of the measurement noise. This set-up is used on carefully prepared samples placed on a compression device to measure the elastic parameters. **Results**: Dentin dilates when the relative humidity increases, with a coefficient of hygroscopic expansion of typically $6.10^{-3}$ %.(%RH)$^{-1}$. This dilatation occurs in about ten minutes. Young modulus and Poisson's ratio are not modified by the variation of relative humidity. **Conclusions**: Hygroscopic expansion is an order of magnitude larger than thermal expansion during dental surgery: around 0.3% with respect to 0.03%. These levels are low with respect to dental rupture, but may induce a significant decrease of the life-expectancy of a restoration.

1. **Introduction**

Dentin is the main tissue of the tooth and thus the support of direct dental restorations. It is a natural composite made of soft and deformable collagen fibrils and strong and rigid hydroxyapatite crystals. Mineral content evolves with age, due to pathologies [1], and is modified during dental restoration especially the superficial layer of dentin which is demineralized prior to bonding of the dental restorative biomaterial.

During usual dental restoration procedures, the mouth and so the dental tissues are under humidity and temperature conditions that may affect the bonding efficiency, especially on dentin [2–5]. The Relative Humidity (RH) has been measured *in vivo* by Plasmans *et al.* [6], and ranges between 75% and 95% in the mouth depending on the geographical location. It can be lowered by the practitioner down to 30% via a rubber dam. Thus, in clinical situation, RH can vary significantly. So, it is crucial to assess the impact of RH on the morphological and mechanical properties of dentin to better understand its impact on the restored tooth properties and further longevity. In particular, it is important to know if dentine inflates with RH, as it may induce significant dimensional variations and thus stresses in the future restored tooth [7]. Similarly, it is important to know the effect of RH on elastic properties of the dentine.

Mechanical properties of dentin, especially its elastic modulus, have been widely studied [8] but a wide range of values is still found in the literature [9]. The values can vary with the inter-patient variability, the patient age, diet etc. but the variability is also often due to the experimental set-ups and conditions. Dentine dimensional changes due to dehydration have been studied through different experimental techniques such as Digital Image Correlation (DIC), Moiré interferometry or Atomic Force Microscopy [7, 10, 11]. It has been proved that a cycle of dehydration/rehydration does not affect the elastic modulus of dentin but its plasticity [12].

However, the studies on the influence of humidity have investigated only the fully hydrated or dehydrated states. In most mechanical researches, there is no humidity control at all. To our best knowledge, no study was led on the sole dentinal substrate at a fixed temperature with known RH variations. To fulfill this gap, we designed an experimental set-up in which the RH and the temperature around the sample are controlled. A first step was to define a special protocol for the preparation of the samples, as perpendicular surfaces are needed for an adequate mechanical assay. Deformations of the dentine were measured by DIC, with a special care on the reduction of the noise.

We measured the coefficient of hygroscopic expansion of the dentine as a function of the RH in a range of 30 to 90%. We also performed compression assays on the same samples, at different relative humidity levels, to measure the evolution of the elastic modulus with the RH. We report a hygroscopic expansion of the dentin, but no variation of the elastic parameter with the RH. This hygroscopic expansion is an order of magnitude higher than the thermal expansion during dental surgery, and may reduce the life of restoration.

2. **Experimental Procedure**
   2.1 **Sample preparation**

Dentin specimens were extracted from sound human third molars. The teeth were obtained following informed consent. The teeth were embedded in epoxy resin. Dentine beams (n=10) were then cut off inside the teeth using a digital milling machine (Lyra-GACD SA, France).

The samples were then polished to have well-perpendicular and parallel surfaces, with an adequate flatness. First, the four long faces of each sample were polished. To do so, the samples were placed in a sample holder, itself placed under a piston device (see Fig. 1). The piston was used to push the holder on the polishing disk, insuring a homogeneous distribution of the pressure on the sample. The polishing was done using SiC polishing papers from 800 to 2000 grit under water irrigation, followed by a final cloth polishing with a 1 µm diamond suspension liquid; the sample was then washed in an ultrasonic cleaner for approximatively 4 minutes to remove the debris. After one face was carefully polished, the sample was turned by a quarter turn, and the next face polished the same way. Once the four large surfaces were prepared, the sample was fixed by applying a small force on a square-shaped sample holder (see Fig. 1). The same polishing procedure was then used for these two end faces.

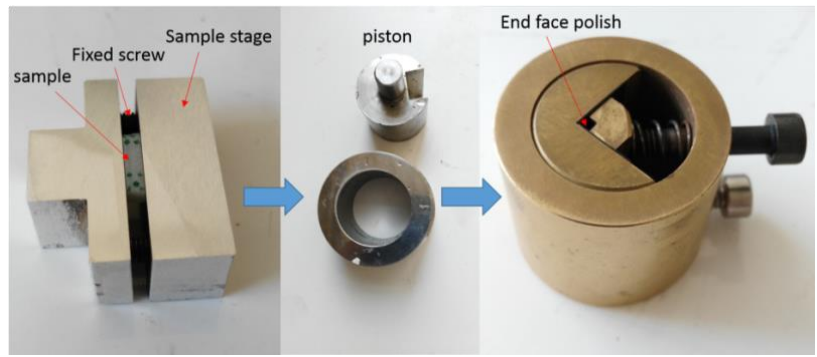

**Fig. 1** Polishing set-ups. From left to right, the sample is placed in a sample holder for the polishing of the large faces (left), and pushed on the polishing paper through a piston (middle). Once the four largest faces have been prepared, the sample is placed into a square-shaped sample holder and polished the same way to make the small surface perpendicular to the large ones (right)

Once cut, the sample was observed using an optical microscope (VHX-1000, Keyence, japan). The angles between faces have to deviate from less than 4 degrees from the expected value (0 for parallel faces, 90 for perpendicular ones), otherwise it was polished completely once again up to meet this criterion. Then the dimensions of all samples were measured (see Table 1). The typical size of a sample is 2.5mm x 1.0mm x 1.0mm (see Fig. 2).

| Sample | Length (mm) | Width (mm) | Thickness (mm) |
|---|---|---|---|
| D1 | 3.05 | 1.16 | 0.95 |
| D2 | 3.06 | 0.77 | 0.66 |
| D3 | 2.58 | 0.79 | 0.65 |
| D4 | 1.99 | 1.00 | 0.88 |
| D5 | 2.88 | 1.36 | 1.28 |
| D6 | 2.19 | 1.07 | 1.05 |
| D7 | 2.47 | 1.10 | 0.78 |
| D8 | 2.14 | 1.14 | 0.95 |
| D9 | 2.02 | 1.42 | 1.40 |
| D10 | 3.03 | 1.00 | 0.90 |
| **Mean** | 2.54 | 1.08 | 0.95 |
| **S.D.** | 0.44 | 0.21 | 0.24 |

**Table 1** Dimensions of the ten dentin samples

An ink speckle was painted on one of the largest surfaces of each sample by using black ink in an airbrush. The quality of the speckle for DIC was checked by visual observations under the optical microscope.

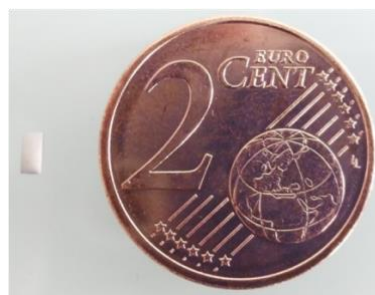

**Figure 2:** Photography of one of our ten samples (sample AC1)

**2.2 Environmental compression set-up**

Compression assays were done using a custom-built compression device, installed in an environmental chamber made of polycarbonate (see Fig. 3a), with a quartz panel for the observation. The atmosphere in the chamber is controlled through an environmental control machine (Secasi technologies, Germany), which regulates the temperature and the relative humidity (RH). A sensor is placed inside of the chamber to measure locally both the temperature and the relative humidity in real time.

The compression device has a classical configuration, with a mobile stage and a fix one (see Fig. 3b). The only specificity is to be built for small samples, with in particular small heads. The force is recorded with a 100 N load cell (Futek, U.S.A.) every 1 s. The displacement velocity of the mobile stage is imposed.

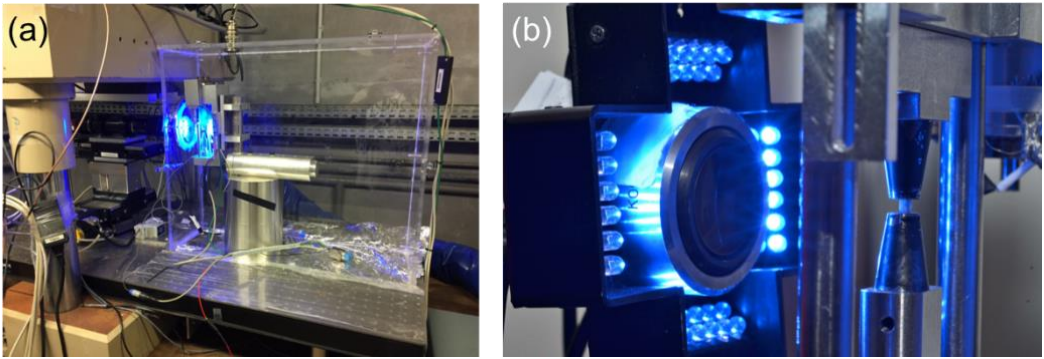

**Fig. 3** Photography of the experimental set-up. (a) global view of the set-up, with the optical microscope mounted on stages on the left and the environmental chamber with the compression device inside on the right. (b) Zoom on the sample in the compression device, with the lightening system around the microscope objective

To measure properly the deformation of the sample, a home-made optical microscope was used to image the speckled surface of the samples at the rate of 1 image per 5 s (0.2 fps) (see Fig. 3a). The microscope is made of a 3296(h)x2472(v) px CCD camera (PROSILICA GX3300, Allied Vision, Germany), connected to an objective lens (x10 Mitutoyo, Japan) with a working distance of 33.5mm and the depth of field of 3.8 µm. A pixel corresponds then to a size of 0.55 µm on the real object. The field of view is 1812X1359 µm$^2$. The whole microscope is mounted on 4 stages: a rotation and three translations (Physik Instrumente, Germany) ones (in X and Y, the directions parallel to the surface with a 3 µm accuracy, and Z the depth normal to the surface with a 0.4 µm accuracy). The rotation stage allows a proper alignment of the optical system with the surface of the sample (so that X and Y translations do not move the surface out of focus). The X-translation moves the camera horizontally, while the Y-translation moves it vertically (the direction of compression). The Z-translation enables to have a proper focus on the surface. To prevent distortions during change in humidity or temperature, the whole microscope was placed outside the environmental chamber, in the air-conditioned room.

The focus was automatically optimized by motions in Z controlled through a contrast maximization algorithm [13]. This enables to preserve the same focus during a whole assay, even if the surface exhibits out-of-plane motions, as due for example to Poisson effect or to non-parallel heads.

Measurement of the surface deformation was done with local Digital Image Correlation (DIC), using home-made software (SYLVIA). DIC measures the local displacement of subsets of the reference image. Each initial subset is characterized by its distribution of grey levels. To determine the new position of this part of the image in a deformed image, the grey-level distribution in the reference image is compared, in the deformed image, to the ones for the different subsets in a vicinity of the previous position of this part of the image. By determining the subset in the deformed image which has the highest correlation of grey-level with the reference subset, the displacement of the center of the subset is determined, as well as the quality (or the confidence) of the correlation [14].

We typically used a mesh of 7x7 squared subsets of 150px, each subset being separated by typically 200px. Once the new positions of the subsets have been determined, the mean deformation of the mesh has been determined through a mean square minimization of the gradient. As our optical method does not allow accessing easily the scale of the tubule, we did not try here to measure locally the strain field. By changing the size of the observation region, we checked that, at the center of the sample, our strain measurement does not depend on the size of the region used for the strain computation (so that we observe an equivalent of a RVE). For this number of subsets, it is possible to have an on-the-fly determination of the mean strain.

### 2.3 DIC Error determination

DIC technique is commonly used and accepted in experimental mechanics [14, 15]. Still, it is hardly used for dental tissues characterization, even if it has proven to be useful [9, 16, 17]. The technique is indeed very attractive since it enables the visualization of the whole strain field instead of the usual average 1D strain assessment on the whole sample length. Despite its strong interest, it is not perfect and we tried to reduce the possible errors. In particular, we installed an autofocus on the imaging set-up to prevent out-of-focus motions and their artificial strains. Errors were quantified to ensure that we were able to measure strains while remaining in the elastic regime of the dentine.

Translation noise:
To determine the noise in the image acquisition and the possible effects of optical distortions, we measured the strain as presented previously for various displacements of the X and Y-translation stages. As translations are rigid-body motions (and as we preserve the focus), we should measure a strain of zero.

A given environment is fixed (T=25°C, RH=55%). The reference image is done in the approximate middle of the sample. Then, one stage moves from -20 µm to 20 µm by steps of 1µm. At each step, an image is taken, and the strain measured.

Out-of-plane error:
If the surface of the sample is closer (resp. further) of the objective than its focal plane, the image will have an artificial dilatation (resp. contraction) which will be proportional to the displacement at least near the focal plane [18].

In the same environment (T=25°C, RH=55%), we moved the Z translation stage from -10 µm to 10 µm by steps of 1 µm, the zero being at the focal plane determined by the auto-focus. At each step, an image is taken and the strain measured.

### 2.4 Relative Humidity dilatation

Dentin dilatation with Relative Humidity was assessed by placing the samples inside of the environmental chamber, without any load. The sample was placed on a support, in front of the camera. The RH was then adjusted at 55%, then 70% and around 90%, followed by another measurement at 70% and finally at 55%. During the whole experiment, the strain at the surface of the sample was measured. At each selected RH, the condition had been kept stable for at least 30 min to let the sample reach its equilibrium. In particular, we checked that no change in strain was observed after this time. The strain was then recorded. The coefficient of hygroscopic expansion for each strain components was then obtained by a linear fit of the strain vs the RH.

On one sample (D6), we measured the coefficient of hygroscopic expansion for very low humidity. The chamber was put to dry to reach 0% RH, and we measured the RH and the strain evolution vs time. On one sample (D7) we measured the coefficient of hygroscopic expansion for RH at 20 and 30%, and on one sample (D9) for RH at 98%.

### 2.5 Elastic parameters

We tested our dentin samples with our environmental compression set-up. The environment was fixed, and we waited long enough so that it was stabilized (typically 60 min). Then, a compression was performed at 1 µm/s (strain rate ranging from $8.10^{-3}\%.s^{-1}$ to $4.10^{-2}\%.s^{-1}$, depending on the sample size and the strain transmission). The compression was stopped at 80N so that the dentin remained in the elastic range. The sample was then unloaded, and the environment changed. On one sample (D6), we tested different compression rate (from 0.1 to 5 µm/s).

The strain was computed by DIC during both load and unloading phases. The stress was determined as the force divided by the initial surface.

The Young modulus from load and unload parts of the sample were determined by linear fits of the two stress vs strain curves of each cycle (see Fig. 4a). The few first points of the load phase may be removed from the fit since they exhibit a strong non-linear response likely due to non-parallel heads.

The Poisson's ratio was calculated as

$$\nu = -\frac{\epsilon_{xx}}{\epsilon_{yy}},$$

and plotted as a function of the strain (see Fig. 4b). The value of the Poisson's ratio was then determined by adjusting a constant, apart from the low-strain non-linear response.

For both Young's modulus and Poisson's ratio, the standard error of the slope was computed to measure the uncertainties on the coefficient.

Sample D5 could not be analyzed with our approach due to the impossibility to obtain a proper fit, the stress vs strain curves being very non-linear although we repeated the experiment and we did not detect any defect in the sample.

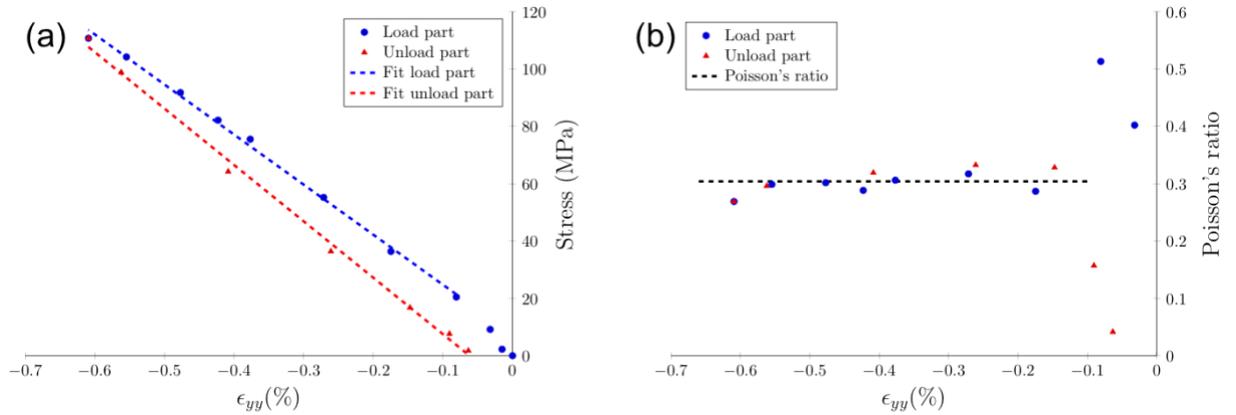

**Fig. 4** Mechanical parameter for sample D8. (a) Stress vs strain curve, with the loading in blue and the unloading in red. Dashed lines represent linear fits, excluding the low-strain non-linear response. (b) Poisson's ratio vs strain, with the loading in blue and the unloading in red. The dashed line represents the fit with a constant, excluding the low-strain non-linear response

In all assays, the temperature was kept almost constant (in the range of 25°C to 30°C, depending on the sample, see Table 2), and we measured the elastic parameters for RH varying from 50% to 90% (as done for the Relative Humidity dilatation). Few samples (n=2, D6 and D7) have been also tested at RH around or lower than 30%, and one sample (D9) has been tested at RH around 98%.

| Sample | Temp. (°C) | 1st RH level (%) | 2nd RH level (%) | 3rd RH level (%) | 4th RH level (%) | 5th RH level (%) |
|---|---|---|---|---|---|---|
| *D1* | 29.8 ± 0.2 | 55.0 ± 0.1 | 72.2 ± 0.2 | 86,6 ± 0.2 | 70.6 ± 0.2 | 54.5 ± 0.2 |
| *D2* | 26.0 ± 1.0 | 56 ± 0.1 | 74.4 ± 0.2 | 92.3 ± 0.3 | 73.7 ± 0.2 | 56.1 ± 0.1 |
| *D3* | 26.2 ± 0.4 | 56.6 ± 0.1 | 75.7 ± 0.1 | 88.7 ± 0.2 | 74.3 ± 0.2 | 55.76 ± 0.2 |
| *D4* | 26.5 ± 0.4 | 55.8 ± 0.1 | 75.9 ± 0.1 | 89.1 ± 0.1 | 75.8 ± 0.2 | 56.3 ± 0.1 |
| *D6* | 29.0 ± 1.5 | 55.8 ± 0.1 | 74.5 ± 0.2 | 89.6 ± 0.2 | 73.7 ± 0.2 | 55.4 ± 0.1 |
| *D7* | 27.1 ± 0.2 | 55.7 ± 0.1 | 75.5 ± 0.2 | 89.4 ± 0.2 | 75.0 ± 0.1 | 55.5 ± 0.1 |
| D8 | 29.2 ± 1.4 | 56.7 ± 0.1 | 75.3 ± 0.3 | 87.9 ± 0.2 | 74.5 ± 0.2 | 54.5 ± 0.1 |
| D9 | 30.7 ± 0.3 | 52.7 ± 0.1 | 66.2 ± 0.1 | 79.7 ± 0.3 | 77.5 ± 0.2 | |
| D10 | 26.6 ± 0.6 | 55.5 ± 0.1 | 74.9 ± 0.2 | 86.4 ± 0.2 | 75.6 ± 0.2 | 55.71 ± 0.1 |

**Table 2** Environmental conditions for the measurement of the elastic parameters of the nine measured dentin samples.

3. Results

### 3.1 Noise determination

We measured the strain tensors induced by translations in X and Y of the camera at RH=20, 33 and 90%. Each component of the strain is in average almost 0, indicating that there is no systematic bias in the measurement. Thus, the standard deviation of the strain is an adequate estimator of the noise. Table 3 summarizes our different measurements. The translation noise does not appear to be correlated with the humidity level. Thus, the errors induced by the changes in the lightening, in the small out-of-plane motion or in the optical distortion during significant displacements of the sample prevent to measure strains smaller than typically $4.10^{-3}$%.

| RH | Translation in X | | | Translation in Y | | | Translation in Z | | |
|---|---|---|---|---|---|---|---|---|---|
| | $\varepsilon_{xx}$ (%) | $\varepsilon_{yy}$ (%) | $\varepsilon_{xy}$ (%) | $\varepsilon_{xx}$ (%) | $\varepsilon_{yy}$ (%) | $\varepsilon_{xy}$ (%) | $\varepsilon_{xx}$ (%) | $\varepsilon_{yy}$ (%) | $\varepsilon_{xy}$ (%) |
| 20% | $1.3\,10^{-3}$ | $9.0\,10^{-3}$ | $3.3\,10^{-3}$ | $2.6\,10^{-3}$ | $2.2\,10^{-3}$ | $2.2\,10^{-3}$ | $8.1\,10^{-3}$ | $1.0\,10^{-3}$ | $3.8\,10^{-3}$ |
| 33% | $2.7\,10^{-3}$ | $2.2\,10^{-3}$ | $2.1\,10^{-3}$ | $2.9\,10^{-3}$ | $3.1\,10^{-3}$ | $2.6\,10^{-3}$ | $8.3\,10^{-3}$ | $7.1\,10^{-3}$ | $7.3\,10^{-3}$ |
| 90% | $4.5\,10^{-3}$ | $2.8\,10^{-3}$ | $1.8\,10^{-3}$ | $3.8\,10^{-3}$ | $3.8\,10^{-3}$ | $3.5\,10^{-3}$ | $1.3\,10^{-2}$ | $5.6\,10^{-3}$ | $3.6\,10^{-3}$ |

**Table 3** Noise values from in-plane translation (X and Y) and out-of-plane translation (Z), at three humidity levels.

Translation in Z (out of the focal plan) of the camera induces an artificial dilatation of the image while remaining centered. During an assay, the sample will move, while the auto-focus will try to replace the surface at the focal plane. Still, it is not perfect, and it guaranties only that the image will be within the depth of field. Thus, by starting from the most contrasted image (associated with the focal plane) and moving in the Z direction by half of the depth of field, we have an estimate of the minimal strain that can be determined, as indicated on Table 3. Thus, an estimate of the error due to the focalization is $1.10^{-2}$%. This error is mainly due to the correlation errors.

Translation errors are of the same order of magnitude, with a slightly higher error due to the out-of-plane motion (see Table 3). Thus, the minimal measurable strain in our experimental set-up is $1.10^{-2}$%, independently of the relative humidity.

### 3.2 Free dilatation due to a change in Relative Humidity

We measured the free dilatation due to a change in Relative Humidity. This was done by changing the Relative Humidity in the environmental chamber while measuring the strains on the sample surface. Figure 5a shows one example of the evolution with time of the RH. While the system needs some time to accommodate the set value, due to external perturbations, the strains follow perfectly the variations in RH. For large variations of RH, a ten-minutes delay can be observed, in agreement with previous measures of the water absorption [11].

In these experiments, the sample is not loaded, so we measured the free dilatation due to the change in RH, similar to the classical coefficient of thermal expansion. The shear strain ($\varepsilon_{xy}$) remains negligible during the whole experiment. The two other strains appear to be proportional to the RH (see Fig. 5b), with slightly different values in the two directions. Table 4 summarizes all our measured coefficient of hygroscopic expansion for the three components of the strain tensor.

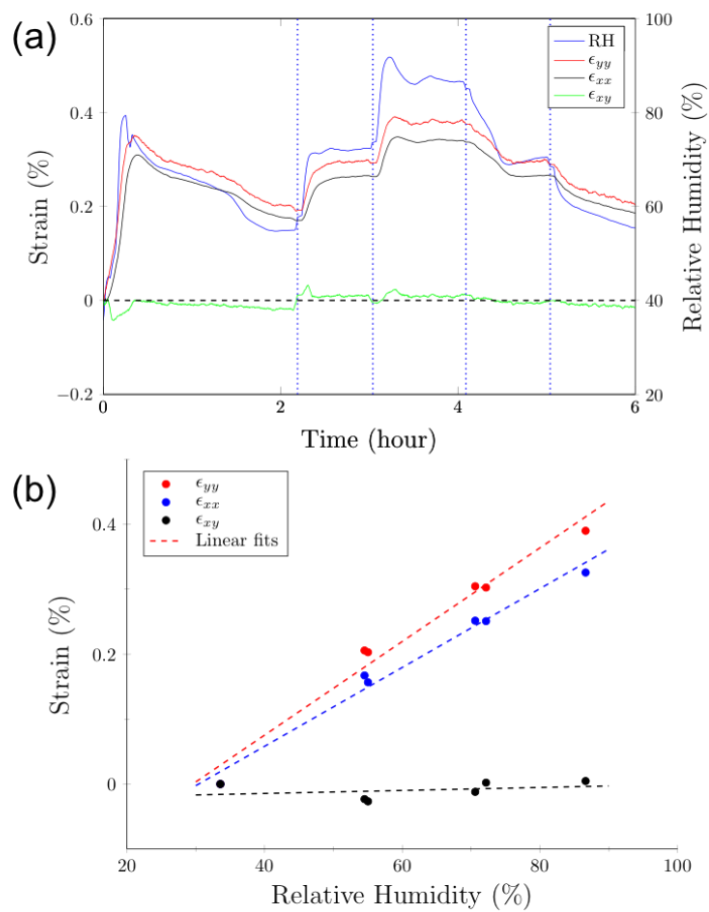

**Fig. 5** Free hydric dilatation of dentin sample D1. (a) Time evolution of the Relative Humidity (RH) and of the associated strains measured on the dentin. (b) Measured strains vs RH, and linear adjustment giving the coefficient of hygroscopic expansion

| Sample | $C_{xx}$ %.(%RH)$^{-1}$ (S.D.) | $C_{yy}$ %.(%RH)$^{-1}$ (S.D.) | $C_{xy}$ %.(%RH)$^{-1}$ (S.D.) |
|---|---|---|---|
| D1 | 6.1 10$^{-3}$ (4 10$^{-4}$) | 7.2 10$^{-3}$ (6 10$^{-4}$) | 2.3 10$^{-4}$ (3.5 10$^{-4}$) |
| D2 | 5.1 10$^{-3}$ (5 10$^{-4}$) | 4.8 10$^{-3}$ (8 10$^{-4}$) | 4.8 10$^{-4}$ (2.1 10$^{-4}$) |
| D3 | 8.2 10$^{-3}$ (8 10$^{-4}$) | 8.7 10$^{-3}$ (1 10$^{-4}$) | -2.9 10$^{-4}$ (6.4 10$^{-4}$) |
| D4 | 5.1 10$^{-3}$ (2.4 10$^{-3}$) | 5.2 10$^{-3}$ (2 10$^{-4}$) | 2.3 10$^{-4}$ (3.5 10$^{-4}$) |
| D5 | 5.8 10$^{-3}$ (3 10$^{-4}$) | 7.3 10$^{-3}$ (4 10$^{-4}$) | 4.3 10$^{-4}$ (1.5 10$^{-4}$) |
| D6 | 4.6 10$^{-3}$ (9 10$^{-4}$) | 6.2 10$^{-3}$ (5 10$^{-4}$) | -7.6 10$^{-4}$ (4.0 10$^{-4}$) |
| D7 | 5.6 10$^{-3}$ (5 10$^{-4}$) | 5.2 10$^{-3}$ (6 10$^{-4}$) | 7.5 10$^{-4}$ (3.2 10$^{-4}$) |
| D8 | 4.6 10$^{-3}$ (4 10$^{-4}$) | 5.8 10$^{-3}$ (3 10$^{-4}$) | -5.6 10$^{-4}$ (5.0 10$^{-4}$) |
| D9 | 5.5 10$^{-3}$ (6 10$^{-4}$) | 5.2 10$^{-3}$ (5 10$^{-4}$) | -3.8 10$^{-3}$ (1.8 10$^{-3}$) |
| D10 | 4.7 10$^{-3}$ (1 10$^{-4}$) | 6.6 10$^{-3}$ (4 10$^{-4}$) | 1.4 10$^{-4}$ (3.1 10$^{-4}$) |
| Mean | 5.5 10$^{-3}$ | 6.2 10$^{-3}$ | 3.2 10$^{-4}$ |
| S.D. | 1.1 10$^{-3}$ | 1.2 10$^{-3}$ | 1.3 10$^{-3}$ |

**Table 4** Coefficients of hygroscopic expansion on our ten samples.

These values were obtained for RH in the range of 30 to 90%. On sample D7, we decreased the RH to 20%, and measured a strain conform to the one expected for a linear hygroscopic expansion. However, on sample D6, we tested a very low relative humidity, 2.5%, and we measured a contraction strongly larger than expected. Starting from a reference state at 30%, we measured $\varepsilon_{xx}$=-5.4 $10^{-1}$%, instead of an expected -1.2 $10^{-1}$%, and $\varepsilon_{yy}$=-8.0 $10^{-1}$%, instead of an expected -1.7 $10^{-1}$%. The sample was not completely at equilibrium after 2.5 hours, so a slightly larger contraction may be expected at equilibrium, reinforcing this tendency. However, as we measured only one sample, this is not sufficient to conclude to a non-linear hygroscopic contraction at low Relative Humidity.

### 3.3. Elastic parameters as function of the relative humidity

At the end of each step of relative humidity, we measured the compressive Young's modulus of the dentin, as well as its Poisson's ratio (see Fig. 4a and b). We limited the stress level to remain in the elastic regime, ensuring that we can continue our experiments on the same sample without alteration: the measured strain never exceeds 1%. At the beginning of the compression, we observed a short non-linear response, typically in the first 0.1%. This non-linear response is likely to be due to the sample accommodation in the loading. After that, the response was in general quite linear, except in the case of sample D5 likely due to a problem of geometry of the sample.

Modules measured in the unload part are generally higher than the ones from the loading part, with a difference of 6% in average going up to 15% (see Table 5). We used the modulus from the unloading part as the Young's modulus of the sample, as it is free from the initial sample accommodation.

| Sample | E (GPa) | $E_{load}$ (GPa) | Relative difference (%) | Poisson's ratio | HR (%) | T (°C) |
|---|---|---|---|---|---|---|
| D1 | 21.1 | 20.9 | -1.1 | 0.31 | 55 | 30 |
| D2 | 21.6 | 21.1 | -2.3 | 0.29 | 56 | 24 |
| D3 | 19.9 | 18.7 | -6.4 | 0.31 | 56 | 26 |
| D4 | 20.6 | 18.9 | -8.4 | 0.36 | 56 | 26 |
| D6 | 12.6 | 11.1 | -11.8 | 0.21 | 55 | 27 |
| D7 | 23.0 | 22.7 | -1.4 | 0.21 | 56 | 27 |
| D8 | 19.7 | 17.5 | -11.2 | 0.31 | 55 | 30 |
| D9 | 6.63 | 6.71 | 1.2 | 0.33 | 53 | 31 |
| D10 | 14.9 | 12.6 | -15.8 | 0.28 | 55 | 26 |
| Mean | 17.8 | 16.7 | -6.3 | 0.29 | 55 | 27 |
| Standard Deviation | 5.35 | 5.36 | NA | 0.05 | 1 | 2 |

**Table 5** Elastic parameters measured for our 9 dentine samples at RH=55% and T=27°C.

Poisson's ratios were very similar in the loading and unloading parts. We observed that it was strongly fluctuating at low strain (when strain was typically less than 0.2%), another evidence of sample accommodation to the compression device at these low loads. We used the mean values, measured outside the initial fluctuation, as the Poisson's ratio of the sample.

Table 5 summarizes all measurements of the modulus and of the Poisson's ratio for the experiments at RH 55% and T 27°C.

We studied the evolution of the elastic modulus with the RH, by measuring the modulus at the 3 RH steps (55%, 75% and 90%, see Table 2). To easily compare the different samples, we plotted the normalized elastic modulus $E_n$ vs the RH (see Fig. 6), $E_n$ being the modulus divided by the mean elastic modulus of the sample. We also added the moduli measured at very large RH (98%) on sample D9, and at low RH (2.5%) on sample D6, as well as the second round of measurements we did on sample D7 on the range of RH from 20% to 90%. We did not observe any significant trend.

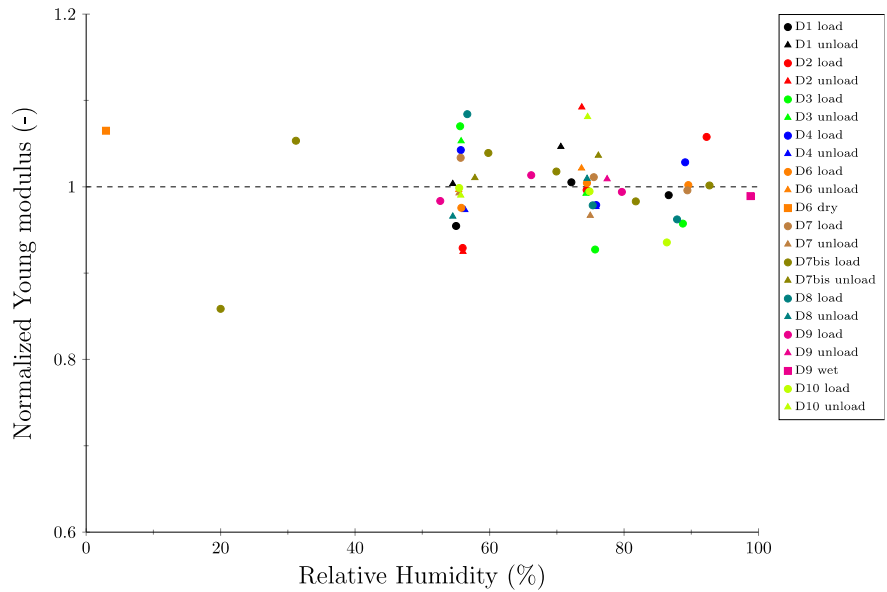

**Fig. 6** Normalized elastic modulus vs Relative Humidity for all dentine samples. Normalized elastic modulus is defined as the modulus divided by the sample mean value of the module

On sample D6, we investigated the effect of the strain rate on the Young's modulus, with strain rate ranging from $8.10^{-4}\%$ to $3.10^{-2}\%$. We observed a weak increase of the Young's modulus with the strain rate, around 1GPa in the whole range in agreement with previous measurements [19].

### 4. Discussion

The free dilatation of ten dentin samples has been studied at different Relative Humidity. The strain is following the variation of RH with time (Fig. 5a), in a reversible fashion [11, 20].

A linear variation of the strain has been observed with RH allowing assessing coefficients of hygroscopic expansion in the two directions (see Fig. 6b). No significant shear has been measured. A slight anisotropy is observed on the average values of the coefficient of hygroscopic expansion (Table 4) with $C_{xx} < C_{yy}$ but the difference between the two coefficients is not in the same direction according to the sample. For three samples out of ten, we have $C_{xx} > C_{yy}$; For six samples out of ten, $C_{xx} < C_{yy}$; And there is no difference for one sample out of ten. We may notice that the larger differences between $C_{xx}$ and $C_{yy}$ ($>10^{-3}$ %.(%RH)$^{-1}$ for D1, D5, D8 and D10, Table 4) are seen when $C_{xx} < C_{yy}$. The slight anisotropy is likely to originate from the local anisotropy in the microstructure of the dentine, with a preferential direction (noted "//") aligned with the tubules and their rigid collar likely to lead to $C_{//} < C_{\perp}$. For very small samples in which the tubules will all be well oriented, this may lead to a strong anisotropy. However, in our samples, the orientation of the tubules changes significantly within the sample (see Fig,7), reducing the anisotropic behavior to an almost isotropic one.

The coefficient of hygroscopic expansion has been found in average to be $5.9.10^{-3}$ %.(%RH)$^{-1}$. To the authors' knowledge this coefficient has not been previously assessed. It implies that the practitioner's choice of protocol may induce significant dilatation of the dentine, with RH ranging from 50% to 100% [21]. Therefore, undesirable stresses due to differential dilatations between dentine and enamel or dentine and a restorative material may appear. A change of RH induces then larger changes than possible thermal fluctuation. Indeed, a thermal variation of 20°C induces a dilatation strain of 0.03% [22, 23] while a change of 50% in RH induces a 0.3% dilatation strain which may induce 30µm displacement for a millimetric sample.

We have also measured the elastic parameters of the dentin at various RH. This was done through a mechanical compression of the dentine well below the yield stress, ensuring that the samples remained in their elastic regimes. The average Young's modulus of the samples is 18.2GPa (±4.9GPa), using our measurements at all RH. We considered the modulus of the unloading part of the assay as the Young's modulus since it is ride of the boundary condition set up. The modulus during the loading part was found to be slightly smaller (17.1 ±4.8GPa). The average Poisson's ratio is 0.31 ±0.06. We used a mean of the loading and unloading part for each sample. Those values are within the (large) range of the measured Young's modulus of dentin under macroscopic compression (1.3GPa-18.3GPa) [8, 9]. They are located in the highest part of this range probably due to the accurate assessment of the strain field using DIC.

We did not observe any dependence of the elastic parameters of the dentine with the RH. Thus, the mechanical parameters do not appear to be coupled with the hydric dilatation. While it is possible that our measurements of the elastic properties are not accurate enough to observe changes with RH, such result is not really surprising for materials in the linear elasticity range.

Forien *et al.* [24] has investigated the molecular consequences of dehydration. They report that the hydroxyapatite minerals compress by typically 0.3%, in agreement with our macroscopic observations. They propose a molecular mechanism based on the contraction of the collagen molecules when the water molecules no longer interact with them. Thus, the collagen resting length decreases, and to preserve the interaction with minerals, it leads to stretched collagen molecules (while still shorter than humid ones) and compressed minerals.

The collagen in dentin is not randomly distributed but presents globally a planar orientation, perpendicular to the tubules [25]. Thus, the hydric dilatation may be expected to be higher in the plan perpendicular to the tubules than in the tubule direction. This explains our observations of slightly anisotropic dilatation, but our samples are too large to have a homogeneous tubule orientation. Tubules in smaller samples, with a typical size of a few tens of micrometers, will be much more aligned and a more anisotropic hydric dilatation will be expected. Another approach will be to measure the strain at a scale small enough to observe the variations between the intertubular region and the highly mineralized peritubular collar, for example under Scanning Electronic Microscopy.

We expect a highest dilatation in the direction perpendicular of the tubules, while it is often considered that the mechanical properties are stiffer in the direction of the tubules. This elastic anisotropy can be explained by the presence of the peritubular collar, highly mineralized, which is surrounding the tubules and so rigidifying dentin in the tubule direction. However, the debate of the dentin Young's modulus dependence on tubule orientation is still open [8, 26–29]. Unfortunately, compression assays cannot measure the anisotropy of the mechanical parameters, contrary to other methods such as RUS [30, 31]. This emphasizes the importance of determining precisely the organization of the tubules networks, but also of smaller structures as the fine branches (and their eventual mineralized collar) [32] to determine their respective contribution in the mechanical and hydric properties.

Our observations show that the hydric dilatation does not modify the elastic properties of the intact dentin. However, in the proposed molecular interpretation, the minerals play a key role in preventing the shrinking of the collagen molecules during drying. Thus, we expect a much higher hydric dilatation coefficient for demineralized dentine, with much drastic consequences if the Relative Humidity is not adequately controlled during surgery.

5. Conclusions

We developed a set-up for the measurements of the mechanical properties of human dentin under different relative humidity. We observed first that dentin expands when the relative humidity increases, leading to significant strains. This expansion does not seem to affect the elastic properties of dentin. This is likely to come from the difference in origin of the two phenomena; Elastic properties originate from the hydroxyapatite crystals, while the hygroscopic expansion comes from a modification in the interaction between water and the collagens molecules around the crystals. The large value of the hygroscopic expansion coefficient is likely to lead to significant residual strains during dental restoration, much larger than thermal effects. This needs to be further investigated, in particular in the context of clinical practice where demineralized dentin is used as a support of dental biomaterials infiltration.

**Acknowledgments**

Authors would like to thanks Vincent de Greef for help in technical implementation. This work has benefited from the financial support of the LabeX LaSIPS (ANR-10-LABX-0040-LaSIPS) managed by the French National Research Agency under the "Investissements d'avenir" program (n°ANR-11-IDEX-0003-02)